\documentclass[12pt]{iopart}
\usepackage{currfile}
\usepackage{graphicx}
\usepackage{amssymb}
\usepackage{amstext}
\usepackage{float}
\usepackage{bm}
\usepackage{epstopdf}
\usepackage{color}
\usepackage{mhequ}
\usepackage{times}
\usepackage{amsthm}
\usepackage{accents}
\usepackage{enumitem}
\usepackage[titletoc,title]{appendix}
\usepackage{perpage} 
\MakePerPage{footnote} 
\usepackage{booktabs}
\usepackage{harvard}
\usepackage{sub_JP}

\def\Label#1{}

\let\phi=\varphi
\let\kappa=\varkappa

\usepackage{chngcntr}
\counterwithin*{equation}{section}

\renewcommand{\theequation}{\thesection.\arabic{equation}}
\renewcommand{\thesection}{\arabic{section}}
\renewcommand{\thesubsection}{\thesection.\arabic{subsection}}

\def\d{{\rm d}}

\let\epsilon=\varepsilon
\let\theta=\vartheta

\let\rho=\varrho

\def\ie{{\it{i.e.}}}

\def\DDR{{\cal D^{\real}\kern-0.8em}}

\def\real{{\bf R}}
\def\cref#1{Corollary~\ref{#1}}

\def\citep#1{\cite{#1}}

\begin{document}
\title{Abelian Sandpiles on Cylinders}
\author{Jean-Pierre Eckmann$^{1,2}$, Tatiana Nagnibeda$^2$, Aymeric
  Perriard$^1$}
\address{$^1$  D\'epartement de Physique Th\'eorique. University of
  Geneva, Switzerland}
\address{$^2$  Section de Math\'ematiques. University of Geneva, Switzerland}

\begin{abstract}
 We study the Abelian Sandpile Model on a special playground, a
 cylinder of width $w$ and of circumference $c$. When 
 $c\ll w$, we describe a phenomenon which has not been observed in
 other geometries: the probability distribution of avalanche sizes has
 a ladder structure, with the first step consisting of avalanches of
 size up to $w\cdot c/2$ that are essentially equiprobable, except for
 a small exponential tail of order about $10c$. We explain this
 phenomenon and describe subsequent steps. 
\end{abstract}

\submitto{\jpa}
\maketitle

  \section{Introduction}

Abelian sandpile models (ASM) come in many guises. Such a model is usually based on a lattice or, more generally, on a connected graph of bounded degree and describes the dynamics of sandpiles on the nodes of the graph. It obeys the following evolution rule: if there are at least $n$  grains of sand on a node with $n$ neighbors, then the node "topples," which means that one grain of sand is moved from this node to each of its $n$ neighbors. It is usually assumed that the system has some dissipative nodes which guarantee that every configuration of sandpiles stabilizes eventually. The beauty of the ASM is that the resulting stable configuration is independent of the order in which unstable nodes were toppled \cite{dhar1990}. The sequence of topplings in the process of stabilizing an unstable configuration is called an avalanche.  
ASM has been invented by \cite{btw1987} to understand the phenomenon of self-organized
criticality (SOC) with emphasis on the power law
behavior of the avalanche distribution.

The ASM on the regular square lattice has been abundantly studied. The 1-dimen\-sion\-al case can be solved exactly and has been proven to be non-critical \cite{dharmajumdar1990} in the sense that the number of avalanches of size $s$ decreases exponentially in $s$. In the 2-dimensional case one approximates the lattice $\mathbb{Z}^2$ by finite squares $\Lambda_n$ with growing size $n$. New grains of sand are added uniformly in $\Lambda$ to re-initiate the avalanches. If grains leave $\Lambda$, they are lost, and this makes the process dissipative.  Various measurements of the avalanches and other correlation functions have been introduced to prove criticality of the model, but it seems that in most cases there is no consensus on the exact power law that they obey (see e.g., \cite{Manna1991,PKI1996}). Besides throwing new grains of sand into the system uniformly at random, another popular choice is to add new grains
of sand always at the same fixed position. One then observes
 beautiful symmetric patterns which form as more and more grains are
 added \cite{propp2010}.

\begin{figure}[ht!]
  \centering{\includegraphics[width=0.6\textwidth]{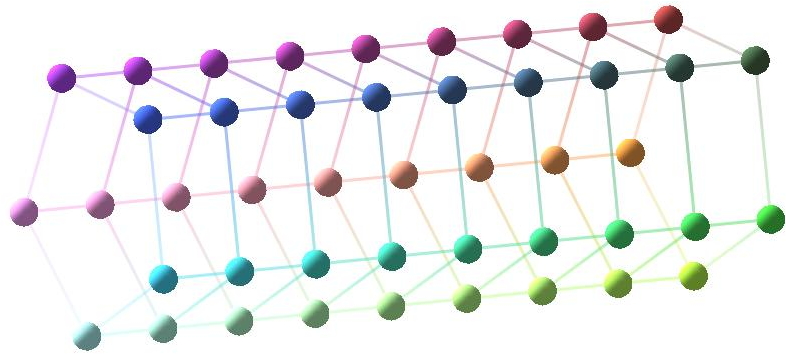}}
  \caption{A cylinder of width $w=9$ and circumference $c=5$. Grains
    can only fall off at the two ends of the cylinder.}\label{fig:cylinder}
\end{figure}

The purpose of our work is to interpolate between the models in 1D and 2D by considering the ASM on discrete cylinders of width $w$ (preferentially an odd integer) and circumference
$c\ge 2$ with dissipative boundary, as shown in \fref{fig:cylinder}.

Regarding the rule according to which the avalanches are reinitialized, we study yet another variant which interpolates
between the two extreme scenarios described above: the grains are added uniformly at random but only near the center of the cylinder,
for example,  uniformly among the nodes with horizontal coordinate $x=w/2 \pm 1 $
and $w/2$ (with $w$ even). We measure the size of avalanches, \ie,  the number of topplings that form an avalanche.

We will study these models for $w$ large, and $c$  small, and will
explain the behavior in a limit of $w\to\infty$, for fixed $c\ge 2$. We will also make some remarks on what happens as $c$ grows and becomes comparable to $w$.

\begin{figure}[ht!]
  \centering{\includegraphics[width=\textwidth]{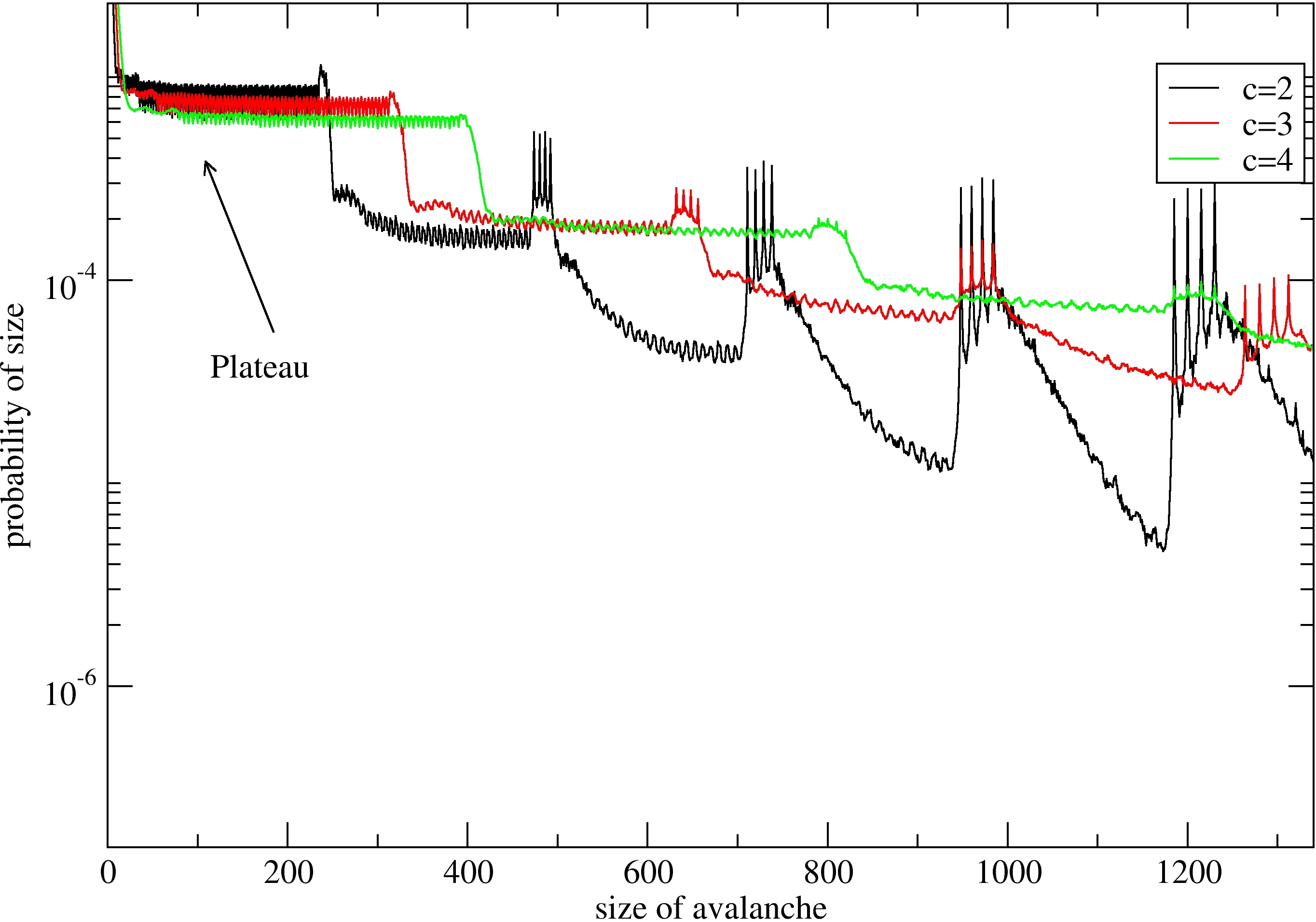}}
  \caption{The probability of an avalanche of size $x$ for cylinders
    of width $w=160$  and circumferences $c=3,4$, and $5$. The first
    plateaus have size about $w\cdot c/2$. The data were acquired by adding  $2\cdot 10^8$ grains (for each fixed $c$). About half of the events produce no avalanche. }\label{fig:steps}
\end{figure}

Our numerical studies show a behavior which seems novel in this context,
see \fref{fig:steps}. Namely, the distribution of avalanche sizes, for
avalanches of medium size, shows  distinct plateaus (steps) of
equiprobable avalanche sizes, which somehow contrasts the idea of
power law decay (it does appear for relatively small avalanches, and
for $c\sim w$,  as discussed in \fref{fig:powerlaw} later on). A
similar behavior was observed in \cite{Daerden1998} who studied the
ASM on a Sierpinski gasket; in this case, the plateaus are connected
with the hierarchical structure of the gasket.

Our main result is an explanation of why these unexpected plateaus
appear, that is, why a large range of avalanche sizes are basically
\emph{equiprobable}. The main observation is that the position of the
end of the avalanche (farthest from the center of the cylinder)
performs a \emph{random walk}. When projected onto $\mathbb Z$,
such random walks visit the sites between the center and the ends of
the cylinder with uniform probability, as is well-known
\cite[``Random Walk on an Interval'', pp.~247--273]{spitzer1964}. Therefore, all sizes of avalanches are seen to be
equiprobable (in the range $0,\dots,w/2\cdot c$).
We will also explain the second and third plateaus
in a similar vein.

\begin{figure}[htbp!]
  \centering{\includegraphics[width=\textwidth]{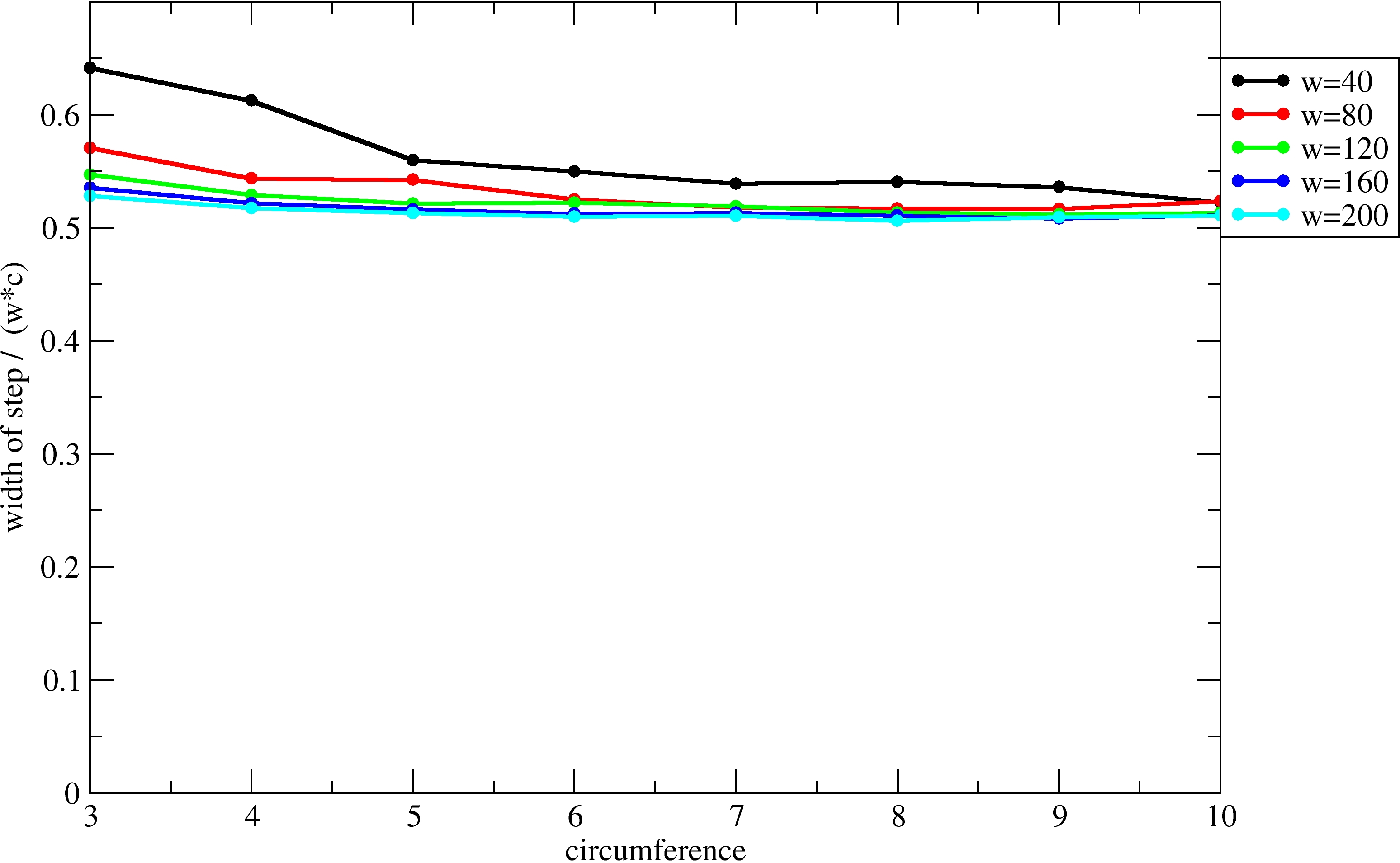}}
  \caption{The width of the (first) plateau, divided by $w\cdot c$ for
    $w=40,80,\dots,200$, and $c=3,\dots, 10$. The convergence to $0.5$
    is manifest.}\label{fig:step1}
\end{figure}
\def\www{0.49}  
\begin{figure}[ht!]
  \centering{\includegraphics[width=\www\textwidth]{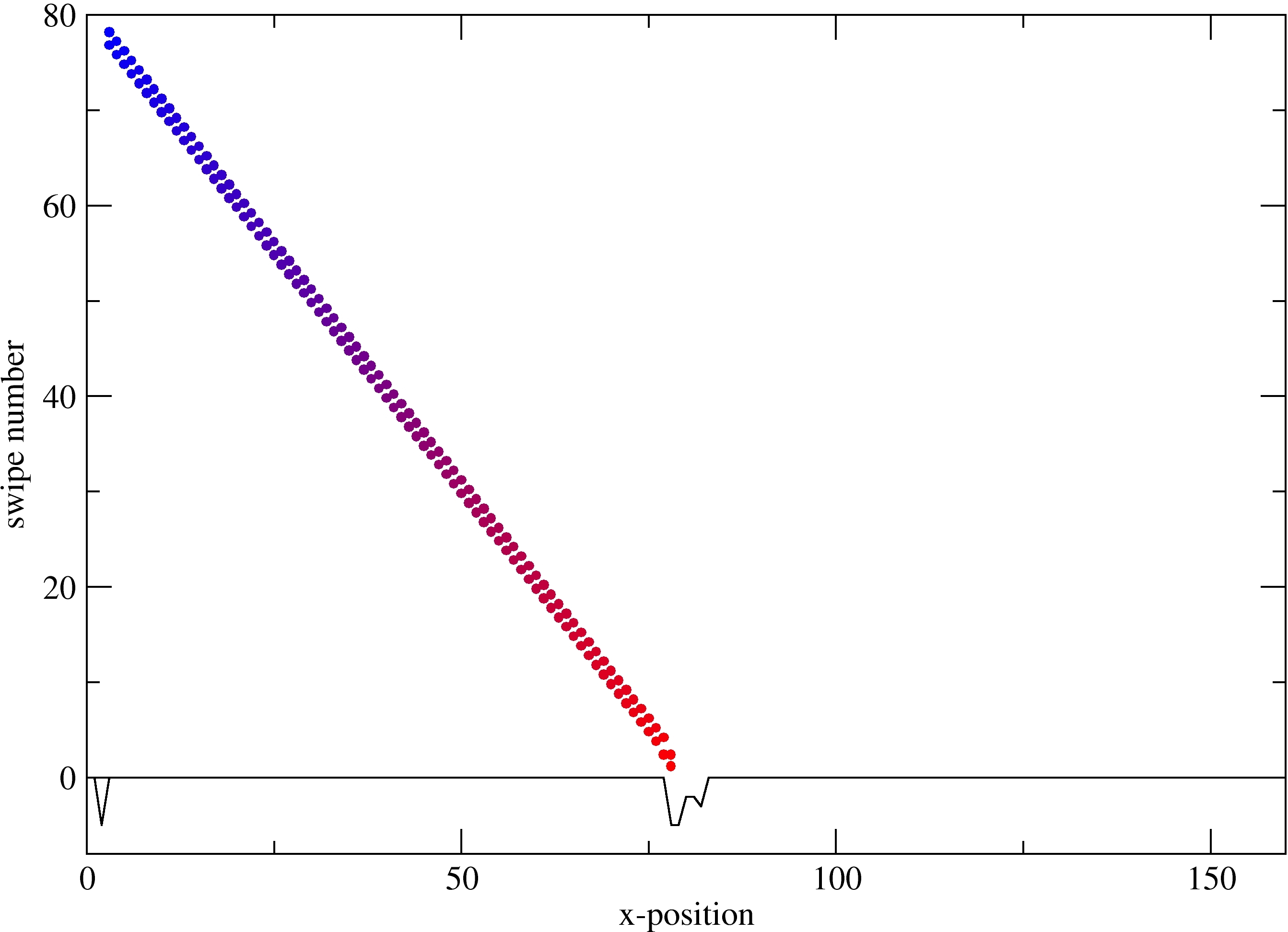}
  \includegraphics[width=\www\textwidth]{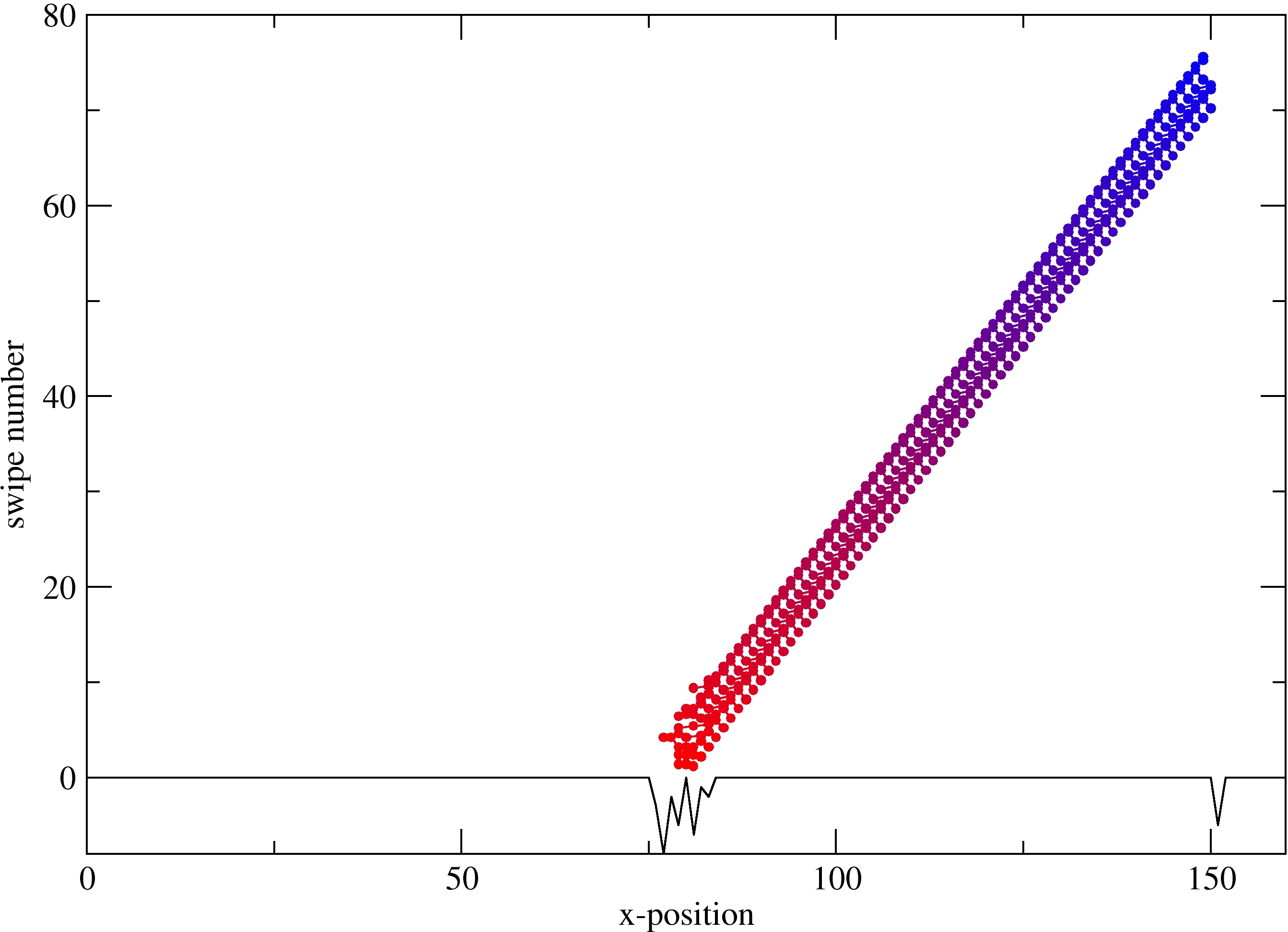}}
  \centering{\includegraphics[width=\www\textwidth]{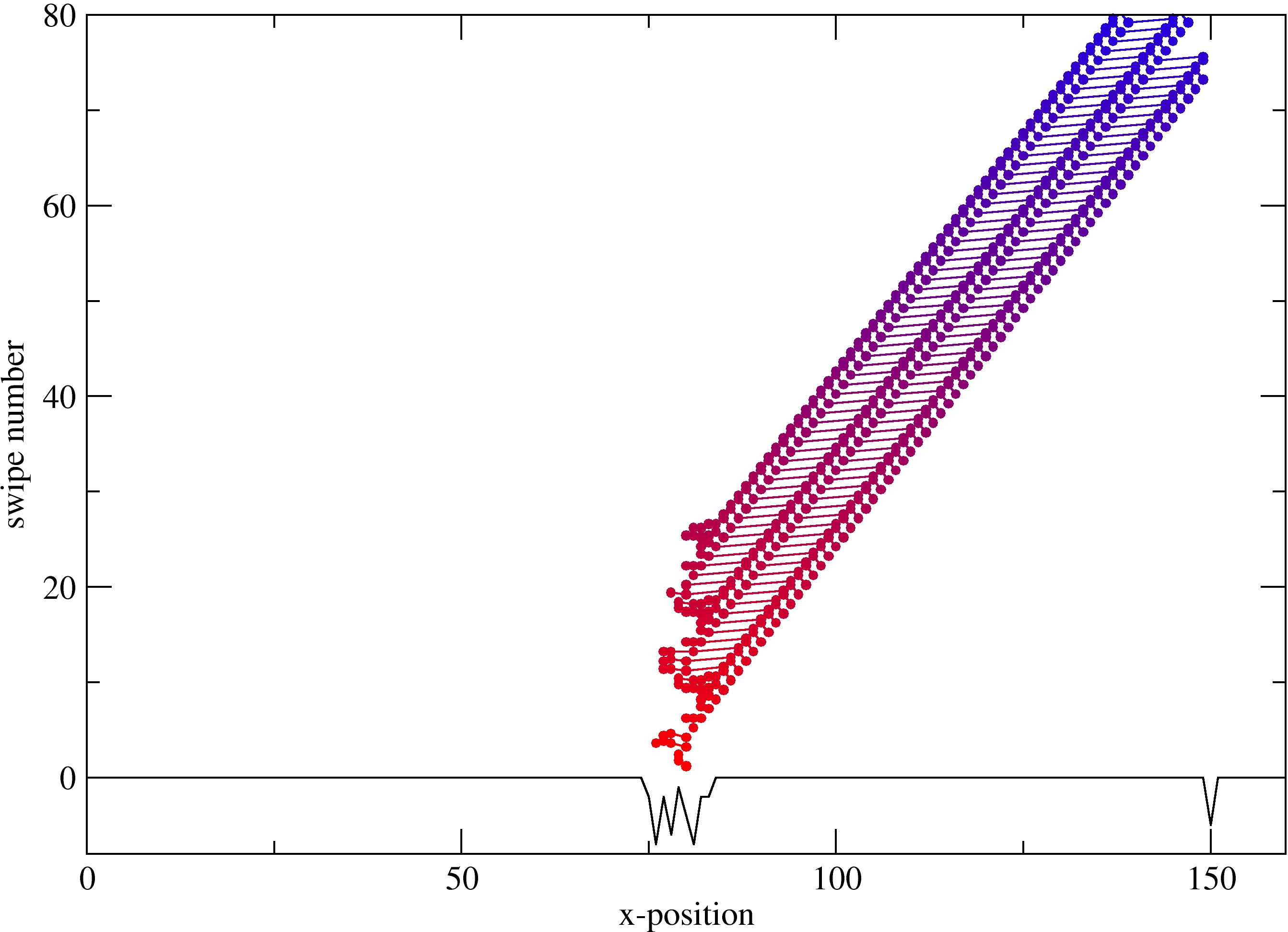}}
  
  \caption{The geometry of a typical avalanche from plateau 1 (top left), 2 (top right), and 3 (bottom). In these samples, $w=160$ and $c=5$. The avalanches have sizes 1516, 2880, and 4300, for the cases of plateau 1, 2, and 3 shown in the 3 panels. The horizontal axis is the $x$ position from $0$ to $w$, the vertical axis is the time steps of the toppling algorithm. These figures illustrate that larger avalanches are created by wider fronts (respectively, single, double and triple for the three panels). These wider fronts occur whenever the random motion in the center creates a big enough set of non-filled positions (as shown by the black line). These black lines at the bottom indicate the filling of the cylinder with the line at height 0 corresponding to all $c$ sites at fixed $x$ having 3 grains. }\label{fig:multi}
\end{figure}

\section{Detailed arguments}

In our model, we are interested in the statistics of the avalanche sizes as a
function of the width $w$ and of the circumference $c$ of the cylinder. Obviously, if the node on which the new grain
is dropped has $\le3$ grains, no toppling happens and no avalanche is formed: we say that the avalanche has size $0$. This happens about $50\%$ of the time.

Inspection of the data shows that, except near the center of the
cylinder, most sites have actually 3 grains (are filled). When a grain arrives, say, from the left on a horizontal position $x\in \{\left \lceil{w/2}\right \rceil,\dots,w\}$, if all $c$ sites in the $x$ position are filled, they will all topple once, and propagate the front of the avalanche to the right. \\


 For each $x\in \{1,\dots,w\}$ we
  compute $s(x)=\sum_{y=1}^c n(x,y)$, where $n(x,y)$ is the number of
  grains at the node $(x,y)$. If $s(x)<3c$ then we say that
  $x$ is \emph{a blocker}. If there is no blocker inside the cylinder on a
  given side from the center, then we  say that the boundary of the cylinder is a blocker (\ie, $x=1$
  or $x=w$).

\begin{figure}[ht!]
  \centering{\includegraphics[width=\textwidth]{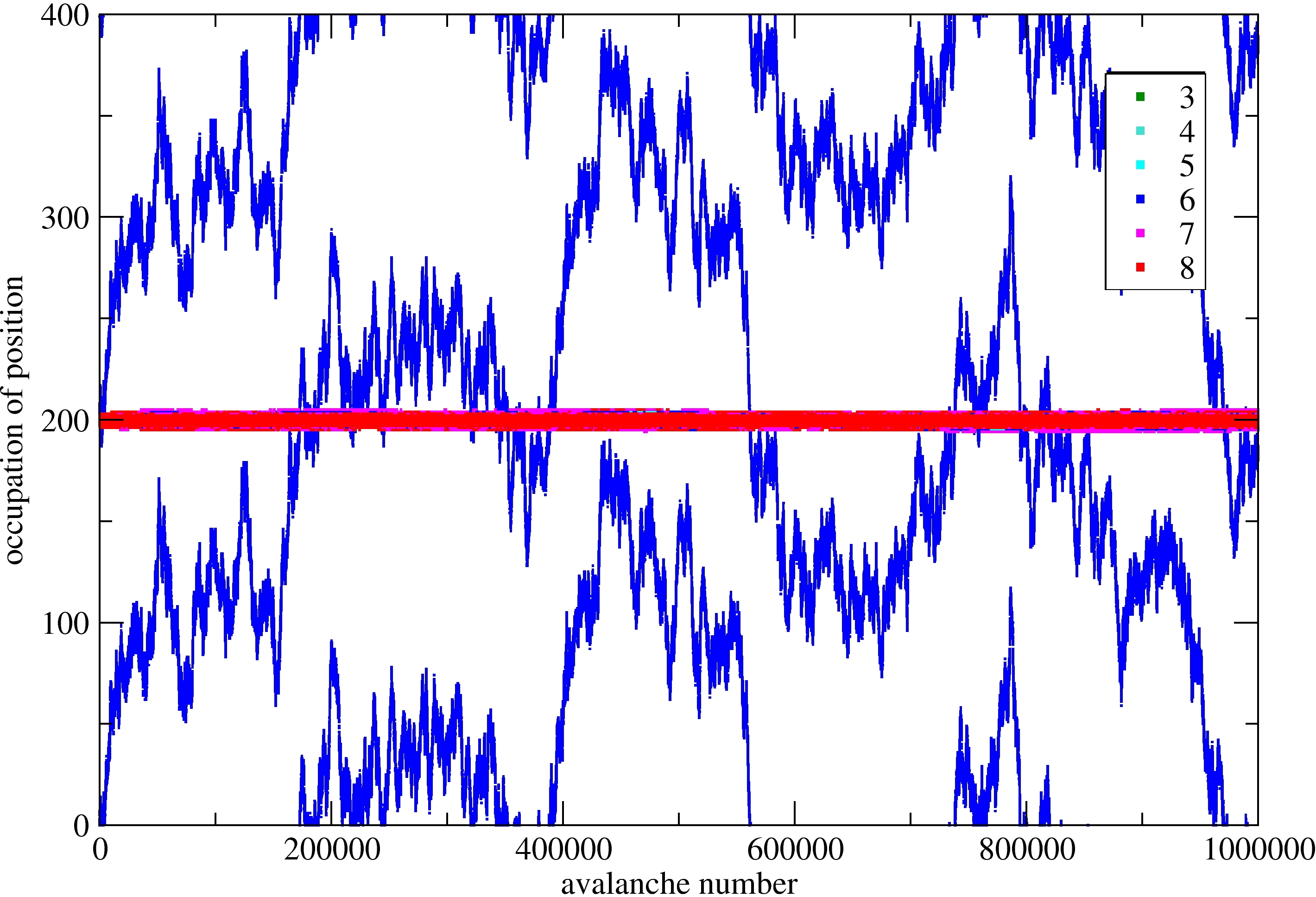}}
  \caption{An illustration of the motion of the blockers (for the first million avalanches) when $w=400$ and $c=3$. 
    The positions of the blocker (coloured in blue) form a trajectory of a random
    walk. Furthermore, the distance between the blockers in the top
    and bottom half of the cylinder (the vertical axis) is always very close  to $w/2$. The blockers (the blue points) always have 6 of the maximal possible number of grains which is $9=3c$. (Actually, 1 grain is missing for every point on the cylinder with fixed $x$.) The red horizontal band near the center, around position 200, is better resolved in 
\fref{fig:blockersdetail} where one also better sees points near the center colored in other colors according to the number of grains (see the color code). Note that "full" sites (with 9 grains) are left white.}\label{fig:blockers2}
   \end{figure}

We start by summarizing the numerical observations, and then we will explain their mathematical origins.\\ 
We first note, as shown in \fref{fig:blockers2}, that there are two
salient features of the model. There is, at any given time, (\ie,
after the system comes to rest when one grain has been added)  exactly 1
blocker, either  on the left or on the right of the center of the
cylinder (we neglect here a region of size about $c$ in the
$x$-direction near the center where more grains may be missing).  
After each avalanche, the blocker either moves by at most $\pm 1$ in the $x$-direction, or jumps to the other side of the cylinder with respect to the center. Inspection of the figure suggests that the distance between the blockers is close to $w/2$, as is seen by the symmetries in the vertical direction. Furthermore, the position of the blocker, as a function of the avalanche count, looks like a random walk in the $x$-direction. 


 Recall that once the system has reached a stable (recurrent)
 configuration the dynamics is reinitialized by adding a new grain
 randomly in the central part of the cylinder. An avalanche is a
 multiset of nodes that become unstable as a result, and we are
 interested in its size. These unstable nodes can be toppled in any
 order resulting in the same new recurrent configuration. As in \cite{PKI1996,dhar2006}, an avalanche is partitioned into ``waves'' in which each mode topples at most once. When examining avalanches on a cylinder, we find it useful to choose moreover a specific order in which the unstable nodes will be toppled. Namely, we perform "swipes" which consist in running
 the cylinder from left to right (or from right to left) and toppling
 the nodes that are unstable. By the Abelian property, the stable configuration in which the avalanche results is independent
 of the order of topplings and in particular of the direction in which the swipe is performed.
     
This allows us to analyze the nature of avalanches in the first 3 plateaus in \fref{fig:multi}, and we see that the avalanche formation is of different qualitative nature: In the first plateau, at each swipe, there are 2 topplings, in the second, 5, and in the third up to 10. These topplings form the advancing front of the avalanche until it reaches a blocker, The front becomes wider as we move to avalanches in the next plateau.\\

\begin{figure}[ht!]
  \centering{\includegraphics[width=\textwidth]{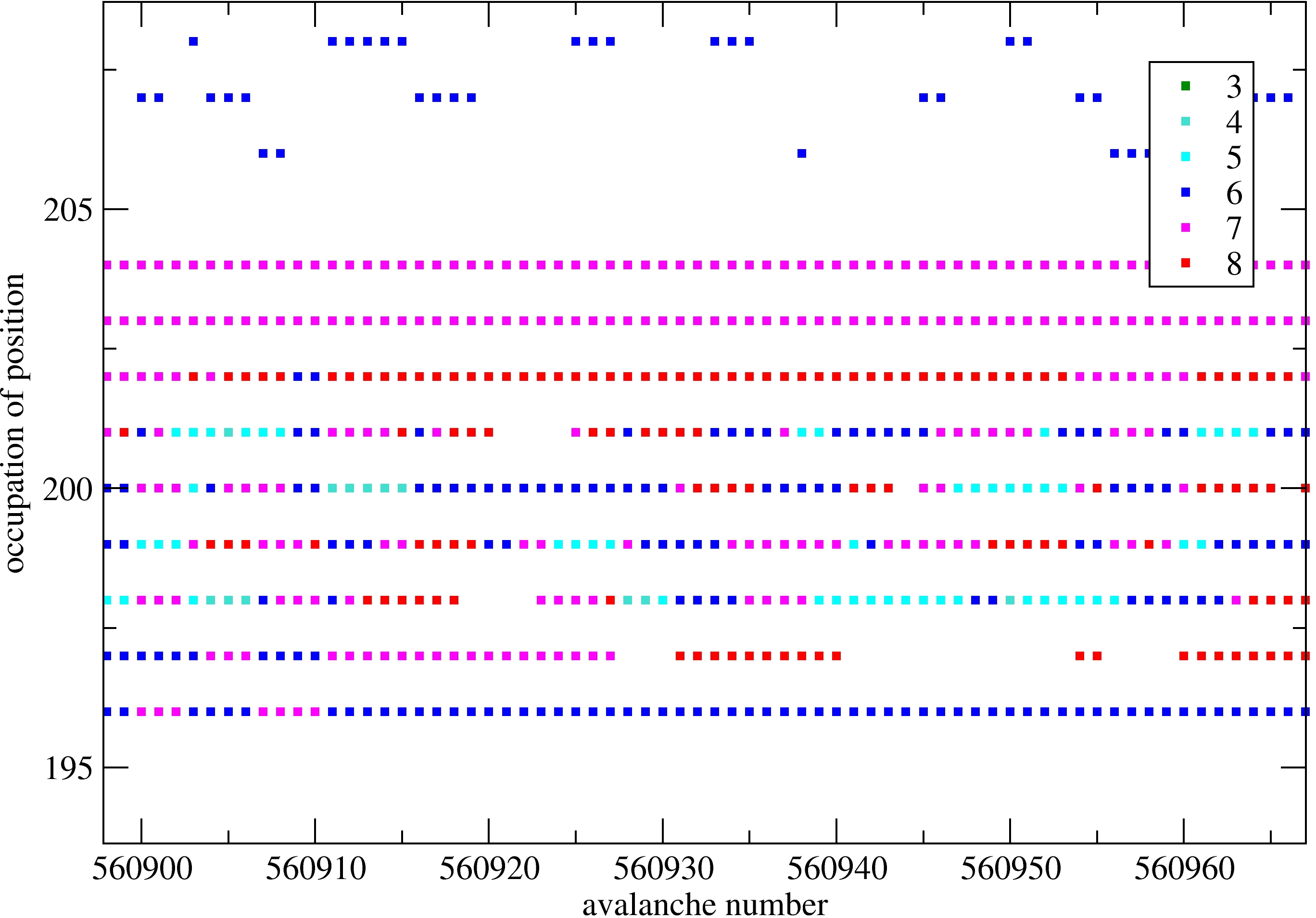}}
  \caption{A magnification of the center of \fref{fig:blockers2}: The red horizontal band near the center is now resolved into a band with a variety of occupation numbers, between 3 and 8, which covers avalanches around 560000 and positions between 194 and 208.  Note that "full" sites (with 9 grains) are left white.}\label{fig:blockersdetail}
   \end{figure}

In \fref{fig:blockersdetail} we illustrate the detailed fillings near the horizontal center of the cylinder. Here, it is visible that there are several sites which are not fully occupied, and bigger holes are the initiators of larger avalanches.

We also check, in \fref{fig:step1} the width of the first plateau (as a function of $w$ and $c$) is basically $w\cdot c/2$.

With these illustrations at hand, we can now prove the existence of the first plateau.
Let $x$ denote the distance between the center of the cylinder and the right blocker. Because the blocker performs a random walk, the density probability function $\rho(x)$ is constant in the interval $d \in (0,w/2)$.\\

When the right blocker is at a horizontal position $x$ and a single front avalanche moves to the right, its size $s$ is equal to $c \cdot x$ (up to some neglected initial topplings in the center).

Thus the probability density function $\rho(s)$ of the single right-moving avalanches is constant, going from $s=0$ to $s=wc/2$ (the biggest avalanche corresponds to $x=w/2$). By symmetry, the case for the left blocker is the same, and thus $\rho(s)$ is independent of $s$.
This corresponds to the first plateau we observe.

\begin{figure}[ht!]
  \centering{\includegraphics[width=\textwidth]{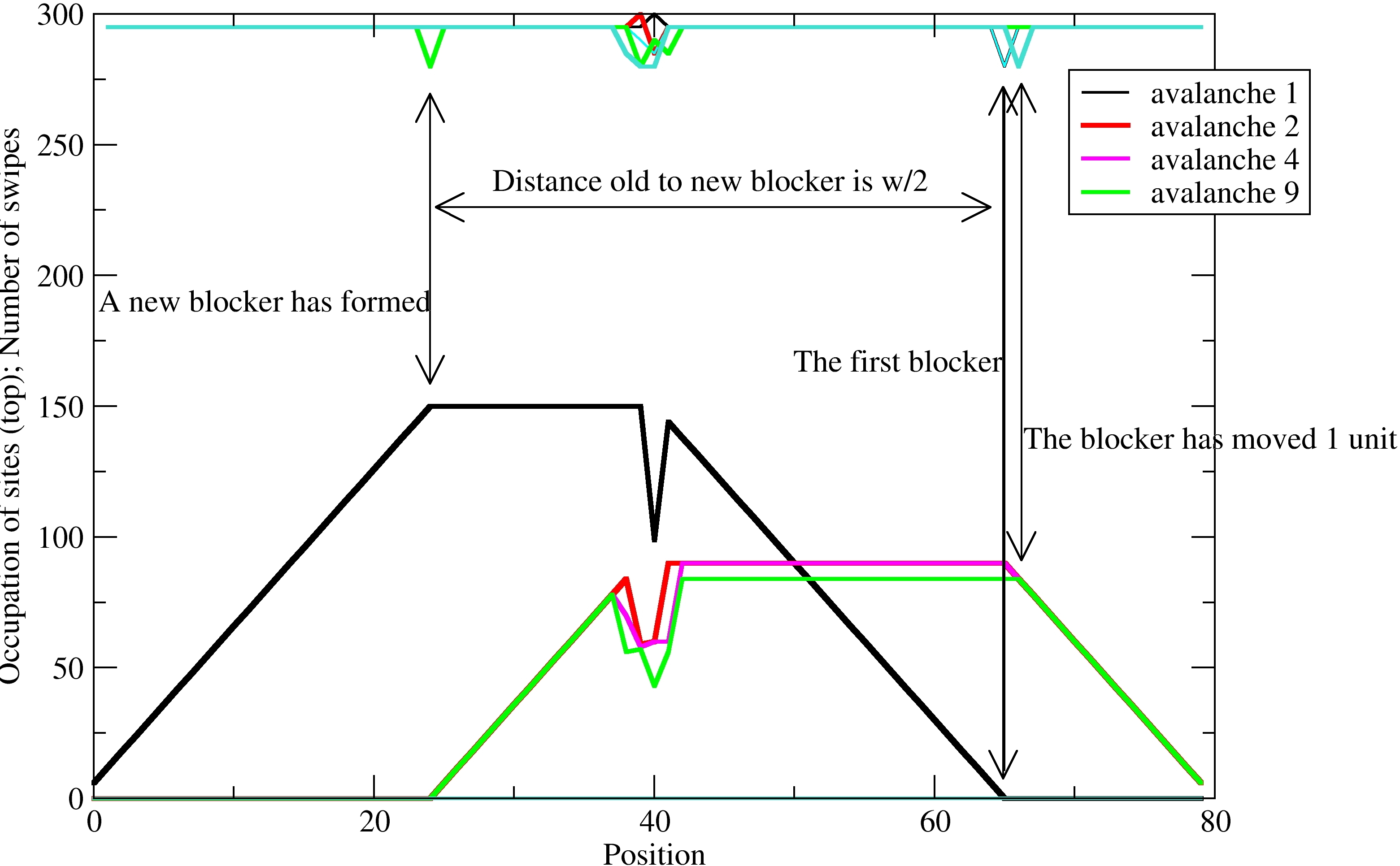}}
  \caption{An illustration of how blockers move. The simulation starts
  with a full cylinder $w=80$, $c=3$ except with one grain of sand
  missing at position 65. We examine several avalanches (others were short, just around position $40=w/2$) that occur when adding one grain to this configuration according to our rules. The first avalanche (in black) needs an increasing number
  of swipes as a function of the distance from 65 and from 0 (the number of the swipe in which the given site will be toppled grows). A new blocker is created  near position 25. In the ninth avalanche (green) the number of the swipes is growing from that position and this entails
  the move of the first blocker one unit to the right. This is how the
  random walk is generated.}\label{fig:blockers}
\end{figure}

\section{Behavior of small avalanches for varying $c$}

\begin{figure}[ht!]
  \centering{\includegraphics[width=0.9\textwidth]{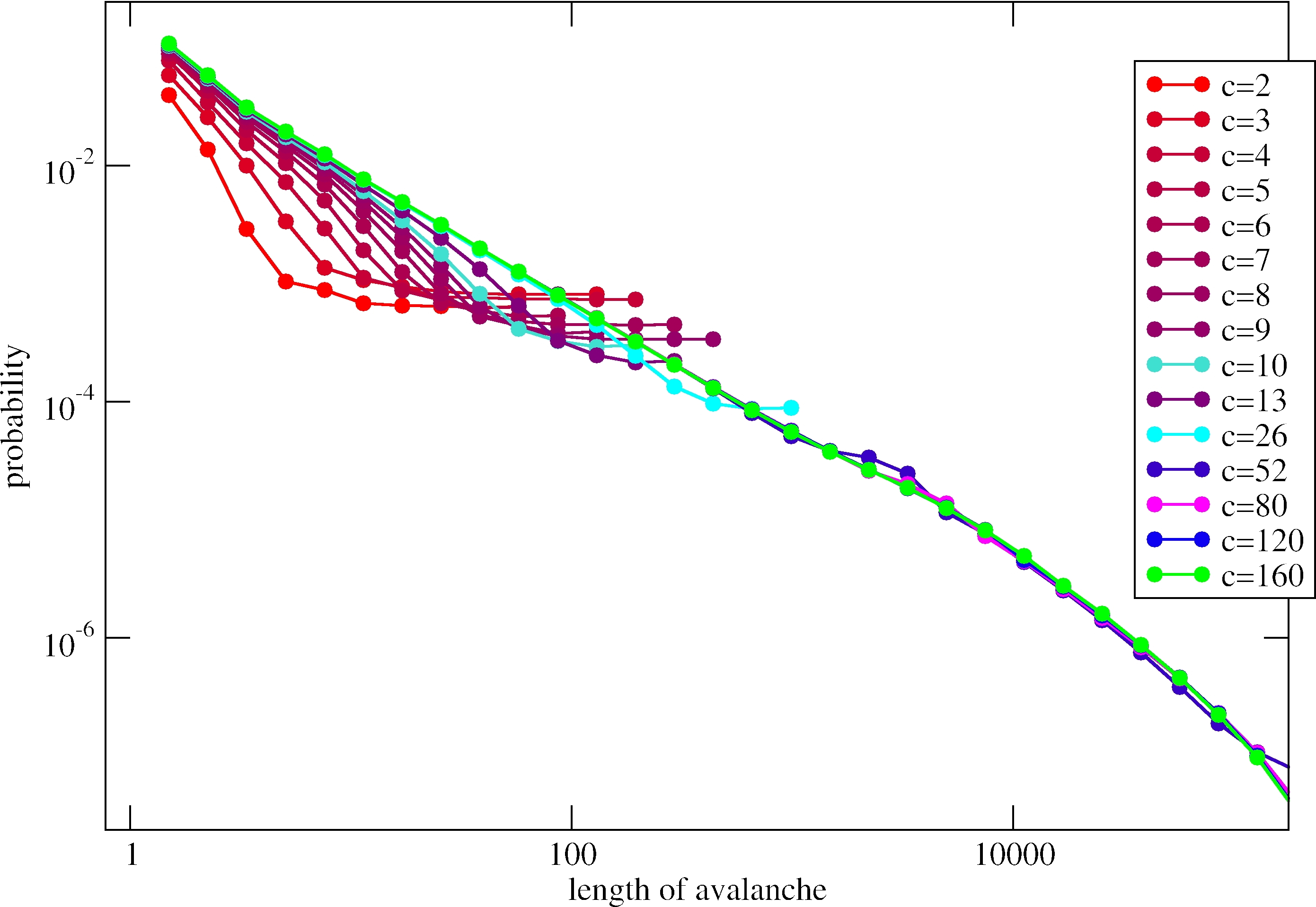}}
  \caption{Power law behavior for small avalanches, for the case
    $w=160$ and several $c$. This is the left end of \fref{fig:steps}, but now the $x$-axis is logarithmic. Note that the
    power law depends on $c$ (for fixed $w$) and reaches $\sim1.035$
    when $c=160$. For small $c$ we see the incipients of the
  first plateau. To make the figure less crowded, for each $c$ we stop drawing the curves once the plateau is reached (the first point where the numerical derivative is positive). The curves are obtained from about $0.85\cdot 10^9$ avalanches.
  }\label{fig:powerlaw}
\end{figure}

We did extensive experiments, with up to $10^8$ rounds of throwing sand, to check the power law behavior, for fixed $w=160$ and varying $c$. The data are sampled logarithmically. Numerical verification shows that for sizes $c=120$ and $c=160$, the decay corresponds to a power law $n^{-\alpha}$, with $\alpha\sim1.035$, see \fref{fig:powerlaw}.
Note also that in this figure, the incipient plateaus are visible for small $c$ and become less and less pronounced as $c$ increases.

\begin{figure}[ht!]
  \centering{\includegraphics[width=0.8\textwidth]{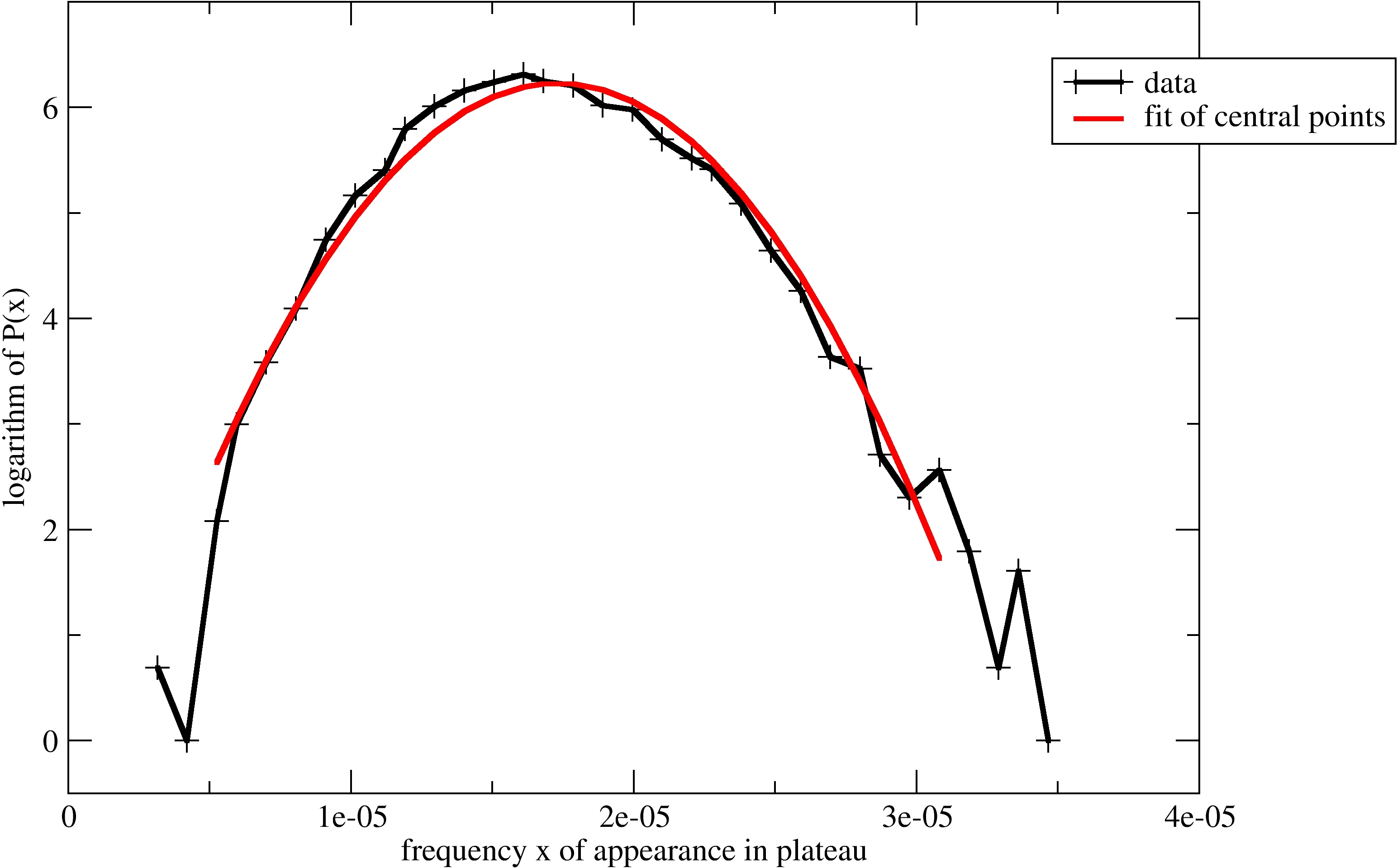}}
  \caption{A numerical verification of the Gaussian fit for $w=4000$
    and $c=3$, for $10^6$ topplings. This makes for a very wide plateau, and we sample the
    number of times each size of the avalanche occurs. This is seen to
  be a Gaussian distribution.}\label{fig:gauss}
\end{figure}

Furthermore, one can consider the avalanches with size between $10c$
and $w\cdot c/2$, \ie, the avalanches forming the first plateau. There are
$n=w\cdot c/2-10c$ possible sizes $s$ of the avalanches, and each one
appears empirically $n(s)$ times when we do $N$ avalanches. We define
$$P(x)\d x=\text{number of }s \text{ for which }n(s)/N\in [x,x+\d x]~.
$$
One expects $P(x)$ to be constant, but there are fluctuations:
Upon about $N\sim 10^6$ topplings for $w=4000$ and
$c=3$, we see that $\log(P(x))$ is close to a quadratic function,
which means that the empirical distribution of $P(n)$ is Gaussian, see \fref{fig:gauss}.
This indicates that the \emph{variations} of the toppling sizes
within the plateau are independent.

\medskip\medskip

\section{The case $c=1$}

It is of course tempting to consider separately the case $c=1$ and  to compare it with the 1-dimensional sandpile models that have appeared in the literature, notably with \cite{alidhar1995,alidhar1995a} where certain decorated chains are studied in great detail. 

In our model, for the case $c=1$, the cylinder is reduced to a chain with a loop attached at each position $x\in[1,\dots,w]$, and the rule we apply is that if there are 3 grains at a site $x$ and one grain arrives to it, the site topples, sending two grains "around" the loop, and one grain to the left and the right,  respectively. In their case, the decorations are either a double link from node $2n$ to $2n+1$ with $n\in\mathbb{Z}$ or a losange inserted at each second node.  The new grains are added randomly and uniformly on every site. A careful analysis of possible avalanche types leads to asymptotic formulas for the avalanche length distribution (in $w$ (their $L$)). In our model we add sand randomly uniformly only to the positions adjacent to the center of the cylinder or to the central positions.  In this case, the avalanches are uniformly distributed over all possible lengths, as can be verified numerically. If one fits a straight, horizontal line,  the standard error is seen to decrease with the size of the system. A an analytic proof seems possible, but is our of the scope of this paper.


\section{$1/f$ behavior}

\begin{figure}[ht!]
  \centering{\includegraphics[width=\textwidth]{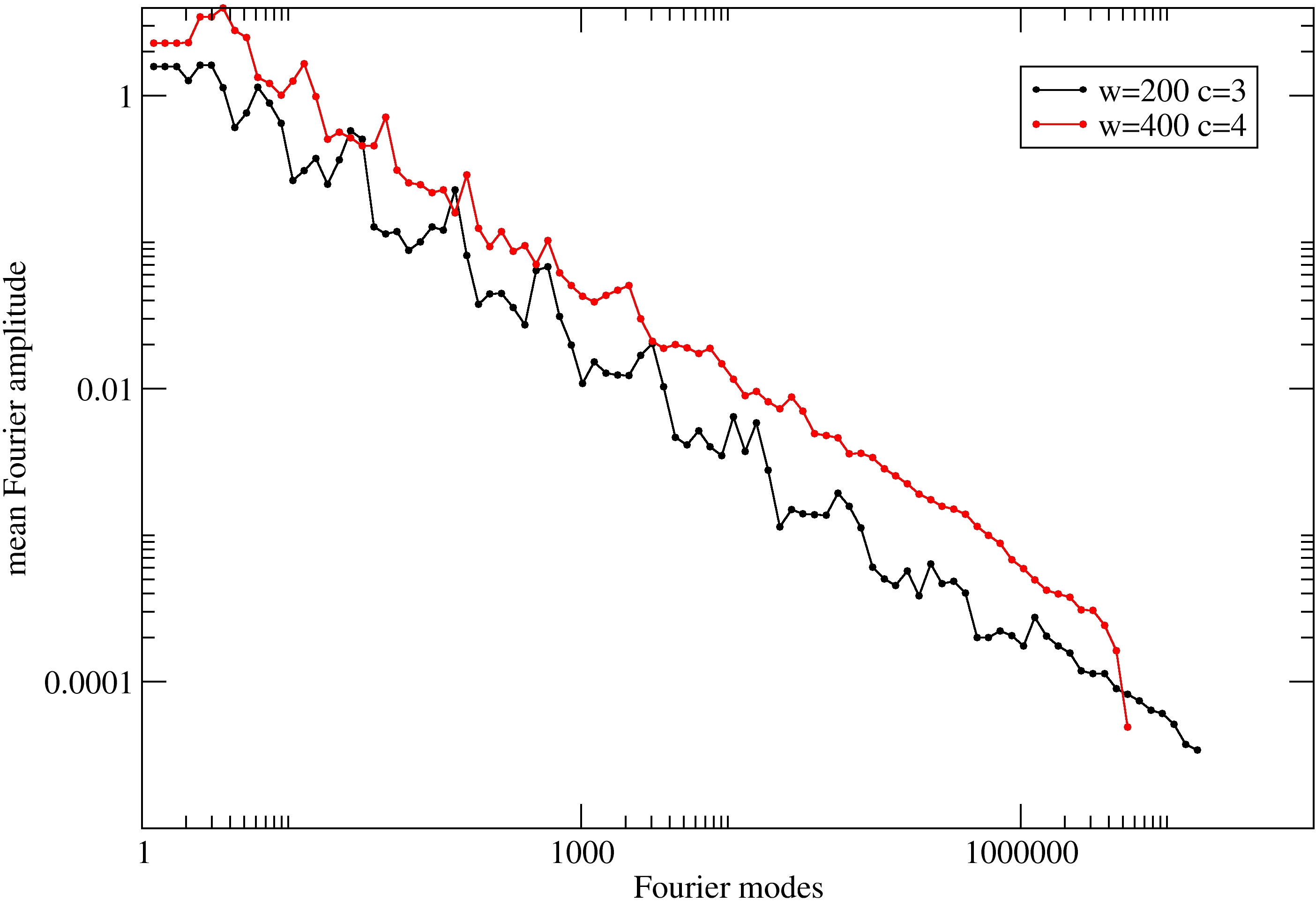}}
  \caption{Let $n(t)$ be the total number of grains in the cylinder
    after $t$ grains have been thrown. We show the Fourier transform
    of $n(t)$ for two values of $w$ and $c$.
    The graph clearly shows a power law (the same in both cases).
     The number of samples is $\sim
    16\cdot 10^6$ for cylinder width $w=200$ and $\sim 5\cdot10^6$ for $w=400$.
  }\label{fig:f}
\end{figure}

Upon suggestion by Ramakrishna Rawaswamy, we also checked the $1/f$
behavior of the filling of the cylinder. The cylinder can have a
maximum of $3w\cdot c$ grains of sand at each site in a stable configuration. We have checked  the
temporal distribution of the number $n(t)$ of grains in the cylinder
after the injection of $t$ grains of sand. We take the Fourier
transform of these numbers (normalized to mean 0) and plot the result
in \fref{fig:f}. The linear fit shows that the power as function of
the frequency $f$ behaves like $1/f^{\alpha}$ for a positive $\alpha$
($\alpha$ depends on the scale of $t$ we take).

\section{Correlations between the process of throwing sand and the motion of the blockers.} 

We have compared our results with the case when the horizontal position where new grains are added is fixed exactly at the center of the cylinder (as opposed to it being chosen randomly over a small neighborhood of the central position as in the previous sections). The avalanche distribution is not the same and we do not observe a random walk of the blocker. 

If the position
where the new grains fall is fixed but not at the center, for example, if $w$ is even, and
the grain falls at $(w-1)/2$, the behavior is as in the random case,
except that the blocker just moves linearly to positions with
lower $x-$coordinates. Therefore, plateaus are visible as in the
random case we studied above. This suggests that there might be a
correlation between the direction of the move of the blocker and the
side at which the grain is randomly thrown, say, at $(w-1)\pm1$ when
$w$ is odd.

\begin{figure}[ht!]
  \centering{\includegraphics[width=\textwidth]{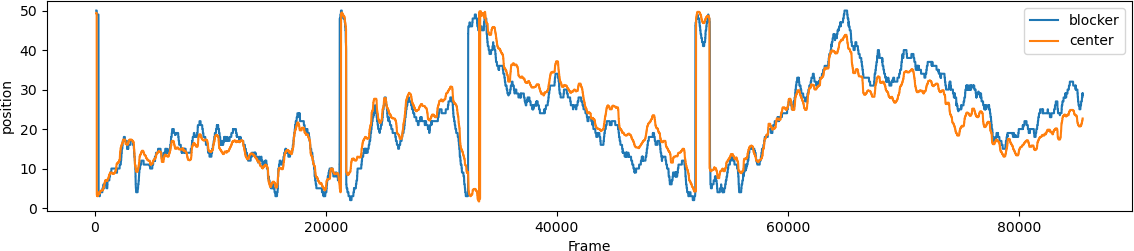}}
  \caption{Correlation between the random walk generated by the $\pm1$
    choices of where the grain is thrown (orange) and the position of
    the blocker (blue); sliding average over 300 data points.
    }\label{fig:pearson}
\end{figure}

In \fref{fig:pearson} we illustrate the random walk for an example
with $w=101$ and $c=6$. While the illustrated data have a Pearson
coefficient of $0.77$, on longer runs there are less good
correlations, because there are additional phase shifts between the
two curves. (The symmetries visible in \fref{fig:blockers2} have been
taken into account by mapping all events to the interval $1\le x\le
w/2$.)

\clearpage

%

%
%
%
%
%


\ack JPE is partially supported by Swissmap. TN acknowledges support of the FNS grant 200020-200400.

\section*{References}
\bibliographystyle{jphysicsB}
\bibliography{sample}
\end{document}